\documentclass[12pt,tightenlines,eqsecnum,floats,aps,amsmath,amssymb,nofootinbib,prd]{revtex4}

\usepackage{setspace}
\usepackage{subfig}
\usepackage{amsmath,amssymb,amsfonts,amsthm,mathrsfs}
\usepackage{graphicx}
\usepackage{enumerate}

\def\be{\begin{equation}}
\def\ee{\end{equation}}
\def\ba{\begin{eqnarray}}
\def\ea{\end{eqnarray}}
\def\bi{\begin{itemize}}
\def\ei{\end{itemize}}

\begin{document}

\title{Gravitation in terms of observables 2:\\ the algebra of
  fundamental observables}

\author{Rodolfo Gambini$^{1}$, 
Saeed Rastgoo$^{2}$, 
Jorge Pullin$^{3}$}
\affiliation {
1. Instituto de F\'{\i}sica, Facultad de Ciencias, 
Igu\'a 4225, esq. Mataojo, 11400 Montevideo, Uruguay. \\
2. 
Department of Physics and Astronomy,
York University,
4700 Keele Street,Toronto, Ontario M3J 1P3, Canada.\\
2. Department of Physics and Astronomy, Louisiana State University,
Baton Rouge, LA 70803-4001, USA.}

\begin{abstract}
  In a previous paper, we showed how to use the techniques of the
  group of loops to formulate the loop approach to gravity proposed by
  Mandelstam in the 1960's. Those techniques allow to  overcome some
  of the difficulties that had been encountered in the earlier treatment. In
  this approach, gravity is formulated entirely in terms of Dirac
  observables without constraints, opening attractive new
  possibilities for quantization. In this paper we discuss the Poisson
  algebra of the resulting Dirac observables, associated with the
  intrinsic components of the Riemann tensor. This provides an
  explicit realization of the non-local algebra of observables for
  gravity that several authors have conjectured. 
\end{abstract}
\maketitle
\tableofcontents{}

\section{Introduction}

In 1962 Mandelstam \cite{mandelstam} published two articles
introducing path dependent techniques for the description of gauge
theories and gravity at the classical and quantum level. The case of
gravity resulted too complicated, but this motivated many physicists
to study gauge theories using related loop techniques. The hope was
that holonomies could allow a better understanding of the confinement
phase.  Makeenko and Migdal \cite{mm}, and Polyakov \cite{polyakov}
proposed different non-perturbative loop techniques in 1979, but the
hopes raised by them were soon abandoned because the very elegant
equations for Wilson loops where difficult to regularize and
renormalize in a non-perturbative way.  In the early 80's, Gambini and
Trias \cite{gatr} introduced the techniques of the group of loops and
the Hamiltonian treatment of Yang Mills in the space of loops. This
technique was subsequently used in Loop Quantum Gravity with an
approach closer to the Hamiltonian formulation of Yang Mills theory.
However, the original idea of Mandelstam was much more powerful.  In
fact, the most ambitious attempt to describe gravity intrinsically
without coordinates and purely in terms of observables was proposed by
Mandelstam in his second paper of 1962. It could serve as the basis
for a coordinate independent approach to the quantization of
gravitation. This paradigm did not flourish because the intrinsic
description loses completely the notion of space-time point, and it
becomes difficult to recover this notion even classically. That is
because in this description, the paths that end in the same physical
point cannot be easily recognized.

Recently, there has been a renewed interest in the description in
terms of observables of gauge theories and gravity. Donnelly and
Giddings \cite{GiDo}  have proposed explicit constructions that
extend the observables associated to gauge theories to the case of
gravitation in the weak field limit. They note that an important feature of
the resulting quantum theory of gravity is the algebra of observables,
which becomes non-local. Observable-based techniques are also used in
several modern developments attempting to extract information from
quantum gauge theories \cite{nonlocal}.

In a previous paper \cite{previous} we have shown how to extend the notion of the
group of loops and its representations arising in gauge theories to
the gravitational case proposed by Mandelstam. This leads to a complete classical description
of gravitation without coordinates. The metric is everywhere referred
to local frames parallel transported starting from a given point. In
such frames it takes the Minkowskian form. The geometrical content of
the theory is completely recovered by relations between reference
frames obtained by parallel transport along paths that differ by an
infinitesimal loop and is given by the Riemann tensor. Although the
construction that we presented there was based on loops, it differs
from the one underlying the usual loop representation of gauge
theories and gravity. In the loop representation the objects
constructed are gauge invariant whereas in the present construction
the objects are both gauge invariant and space-time diffeomorphism
invariant. That is, the objects are Dirac observables. This leads to a
theory that does not involve diffeomorphisms and may allow to bypass
at the quantum level the LOST-F \cite{lostf} theorem that leads to a discrete
structure in the Hilbert space of ordinary loop quantum gravity and
conflicts with the differentiability of the group of loops. The latter
is crucial to recover the kinematics of gauge theories and gravity in
this context.

In this paper we will show how to determine Poisson brackets among
path dependent Riemann observables that are consistent with Einstein's
equations. In his original papers Mandelstam had computed an algebra
of Poisson brackets but it was unclear whether it was compatible with
the Einstein equations. It turns out that the resulting algebra among
Riemann observables evaluated on arbitrary paths is
non-local. Observables do not organize themselves into local commuting
sub-algebras (as occurs in usual field theories), and therefore the
principle of locality \cite{haag} must apparently be reformulated or
abandoned, as Donnelly and Giddings \cite{GiDo} conjectured.  In
section II we review the Mandelstam intrinsic formulation and the
techniques allowing to determine physical points. In section III we
determine the Poisson algebra of path dependent Riemann tensors in
vacuum, finally in section IV we concluded with a discussion of the
non-locality of the algebra of the gravitational theory and some final
remarks.

\section{Previous results}

Mandelstam's construction starts by the {\em intrinsic} specification
of paths in a manifold. By that he means the following: starting from a chosen initial point (in
asymptotically flat manifolds it could be infinity) one parallel
transports a frame along a curve a certain invariant distance and then follows
another distance along a different direction and so on. The important
point is that the direction is defined by the parallel transported
frame. Therefore one  characterizes curves by a series of instructions
of how to proceed with respect to a local frame. This has similarities
with how a GPS provides instructions to a driver to follow a path. 
Diffeomorphisms in space-time affect the curves but not the set of
instructions that is given intrinsically. The trouble with these sets
of instructions is that it is difficult to determine if two paths end
at the same point. This hampered the development of Mandelstam's
framework in the 1960's. 

Suppose one considers two paths that intrinsically are sets of 
instructions opposite of each other. Clearly, if one were to follow them,
one would return to the same point. Suppose, however, that one adds an
infinitesimal loop between them. The loop would alter the frame with
respect to which the initial instruction of the return path is
specified. As a consequence, one would end up with a path that does not
return to the same point, as shown in figure 1.
\begin{figure} 
\includegraphics[height=6cm]{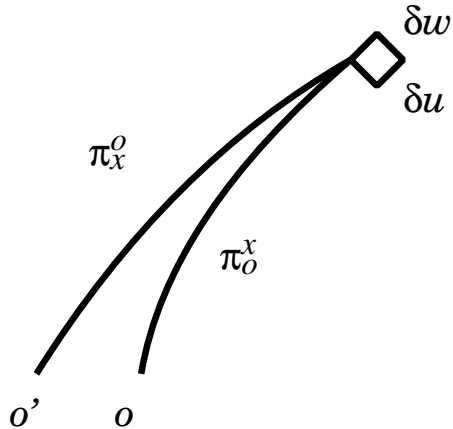}
\caption{The path has the same intrinsic initial and final coordinates (that is why both paths are
  labeled by $o$ and $x$) but would correspond to two
  different end points of the space-time, $o$ and $o'$.}
\label{condef} 
\end{figure} 

If one wished the path to start and end at the same point, one would
have to correct the set of instructions of the return path to undo the
rotation of the frame that took place due to the addition of the
infinitesimal loop. This way, if one considers the frame one started with and
evaluates the parallel transport of it along the corrected path, one
gets back to the same point. The initial and final frame of such a
closed path would be related by a Lorentz transformation (holonomy)
given by,
\begin{equation}
H(\pi_o^x\circ \delta{\gamma}\circ \Lambda(\delta{\gamma})\pi_x^o)^\alpha{}_\beta=
\delta^{\alpha}{}_\beta + \delta u^\rho \delta  w^\sigma R_{\rho
  \sigma}{}^{\alpha}{}_\beta(\pi_o^x),
\end{equation}
where $\Lambda(\delta \gamma)\pi_x^o$ is the retraced rotated path
described above and $R$ is the Riemann tensor. 

The addition of an infinitesimal loop is associated with the generator of the group of
loops. The composition (product) of such infinitesimal generators can
be used to construct finite loops. This allows to reverse the
construction: two intrinsically defined paths will end at the same
point if they differ by a loop. This was the missing piece in
Mandelstam's 1960's construction that we added.

In this framework, matter fields become path dependent, and they are
given by the fields evaluated at the endpoints of paths. Under a
change of path, they transform with appropriate holonomies. 
 For instance, for a vector field
with internal group $SU(N)$ in some representation,
\begin{equation}
A^\alpha{}_I(\pi') = H(\gamma)^\alpha{}_\beta H\left(\gamma\right)
_I{}^J A^\beta{}_J(\pi),
\end{equation}
if $\pi'=\gamma\circ\Lambda(\gamma)\pi=\gamma\cdot \pi$,
which guarantees that $\pi'$
and $\pi$ end at the same point on $M$. Here, we introduced the dot as a
shorthand for the composition of loops in the intrinsic formulation,
incorporating the Lorentz rotation of the previous loop. Also, $H(\gamma)^\alpha{}_\beta$ is a
holonomy associated with the Lorentz group and $H(\gamma)_I{}^J$ a
holonomy associated with the internal group.

In our previous paper we saw that it is possible to identify when
two paths described in such a way end at the same point. Indeed, for
a given geometry, two open paths $\pi,\pi'$ whose local bases
transported to their ends differ by a Lorentz transformation and
$\pi'=\gamma\cdot \pi$ with $\gamma$, then both paths end at  the same
point in the manifold $M$. The notion of closed loops also depends on
the geometry. This implies that at the quantum level, when the
geometry fluctuates, so do the points, and they become fuzzy objects. 

The intrinsic quantization is therefore {\em nonequivalent} to the
usual one. Let us be more explicit using the technique developed in
section VII of our previous paper \cite{previous}. Given a path $\gamma^a(\lambda)$
in a differential manifold in a given coordinate system,
with $\gamma^a(0)=x_o^a$ the coordinates of $o$ and $\gamma^a(1)=x^a$
a local point, the frame transported along $\gamma^a$ is,
\begin{equation}
  e_\alpha{}^c(\lambda)={\rm P}\left(\exp\left(-\int_0^\lambda
      d\lambda'\dot{\gamma}^a(\lambda') \Gamma_a\right)\right)_d^{\,\,\,c} e_\alpha{}^d(0),  
\end{equation}
where $\Gamma_a$ is the connection in the given coordinate system, 
and the intrinsic coordinates are given by
\begin{equation}
  \label{eq:1}
  y^\alpha(\lambda)= \int_0^\lambda \dot{\gamma}^c(\lambda')
                     e^\alpha{}_c(\Gamma,\lambda') d\lambda',
\end{equation}
which implies that upon quantization of the geometry (and therefore of
$\Gamma$), to a curve
$\gamma^a(\lambda)$ in $M$ corresponds an operator
$\hat{y}^\alpha(\lambda)$. On the contrary, if one considers intrinsic
coordinates as the primary description of the path and one uses,
\begin{equation}
  \gamma^a(\lambda)= \int_0^\lambda d\lambda' \dot{y}^\alpha\label{7.17}
                     e_\alpha{}^a\left([y],\lambda'\right)+x_o^a,
\end{equation}
where $e_\alpha{}^a\left([y],\lambda'\right)$ is the tetrad
transported from the origin with the prescription given by the
function $y$ up to the point with parameter $\lambda'$,
one would get for the intrinsic trajectory $\gamma(\lambda)$, upon quantization, an
operator,
\begin{equation}
  \label{eq:3}
  \hat{\gamma}^a(\lambda) = \int_0^\lambda d\lambda'
  \hat{e}_\alpha^a\left([y],\lambda'\right) \dot{y}^\alpha +x_o^a,
\end{equation}
and therefore the traditional notion of curve only is recovered in the
semi-classical approximation. In the usual quantization scheme one is
given a curve that remains classical and quantizes the geometric
operators, like the metric. Intrinsically defined curves become quantum operators, as
they depend on the metric. Conversely, 
if one were to take the intrinsic description of the curve as a starting point for a quantization,
the latter would be classical whereas the curve itself becomes a
quantum operator as shown in (\ref{eq:3}). 

\section{Procedure for computing the Poisson brackets for intrinsic
  components of the Riemann tensor}

If one takes as reference paths in the action presented in section IX
of the companion paper the ones used in Riemann or Fermi normal
coordinates, one recovers the standard Einstein--Hilbert action in
those coordinates. It is well known \cite{Nesterov} that in order to
have geodesics that do not cross each other, one must have
$s\ll \vert R_o\vert^{-1/2}$ where $R_o$ is the typical size of the curvature
and $s$ is the length of the geodesic. In this region of validity we
can use the Palatini first order action,
\begin{equation}
  \label{eq:11}
  S = \int d^4x \sqrt{-g} g^{ab} R_{ab}\left(\Gamma\right),
\end{equation}
and recalling that in normal coordinates $g_{ab}=\eta_{ab}+h_{ab}$
with $h$ of second order (in $s$) and $\Gamma_{ab}^c$ is an independent first
order quantity and the $R_{ab}$'s are zeroth order quantities, 
\begin{equation}
  \label{eq:12}
  R_{ab} =\Gamma_{ab,c}^c-\Gamma_{ac,b}^c +\Gamma_{ab}^c \Gamma_{cd}^d
  -\Gamma_{ac}^d \Gamma_{bd}^c.
\end{equation}

Recall that either in Riemann or Fermi normal coordinates the gauge is
partially fixed. This will not be relevant because we are going to
compute the relation for diffeomorphism-invariant quantities at the
end, and working in a specific coordinate system simplifies the
canonical analysis.

Let us restrict the action considering its expansion up to second
order in $s$. The analysis is valid for arbitrary Riemann tensors in
a sufficiently small region. The action will then read,
\begin{equation}
  \label{eq:13}
  S_2 =\int d^4x \sqrt{1+h_d{}^d}\left(\eta^{ab} -h^{ab}\right) R_{ab}(\Gamma),
\end{equation}
where $h$ and $\Gamma$ are considered independent variables. The
variation with respect to $h$ yields $R_{ab}-\eta_{ab} R/2$ with
$R=\eta^{cd} R_{cd}$. Variation with respect to $\Gamma$ leads to,
\begin{equation}
  \label{eq:14}
  \left(\sqrt{-g}g^{ab}\right)_{;c}
  =\left(\sqrt{1+h_d{}^d}\left(\eta^{ab}-h^{ab}\right)\right)_{,c}
  + \eta^{ad} \Gamma_{dc}^b +\eta^{bd} \Gamma_{dc}^a
  -\eta^{ab} \Gamma_{dc}^d,
\end{equation}
which implies that the first order connection takes the form,
\begin{equation}
  \label{eq:15}
  {}^{(1)}\Gamma_{ab}^c  =\frac{1}{2} \eta^{cd}
  \left(h_{ad,b}+h_{bd,a}-h_{ab,d}\right).
\end{equation}

Introducing adapted three dimensional quantities,
\begin{eqnarray}
  \label{eq:16}
  g_{ij}&=& \eta_{ij} +h_{ij}, \qquad N=\left(-\eta^{00}
            +h^{00}\right)^{-1/2}=1-\frac{h^{00}}{2},\nonumber\\
N_i&=& h_{0i},\qquad \sqrt{g} =\sqrt{1+h_i{}^i},\qquad \sqrt{-{}^4g}=N\sqrt{g}\nonumber\\
\pi^{ij} &=&\left(\Gamma_{pq}^0 -\eta_{pq} \eta^{rs}
  \Gamma_{rs}^0\right) \eta^{pi} \eta^{qj} =
  \Gamma_{pq}^0 \eta^{ip} \eta^{jq} -\Gamma_{rs}^0
  \eta^{rs} \eta^{ij},\\
\Gamma_{pq}^0 &=&\frac{1}{2} \left(h_{pq,0}-N_{p,q}
  -N_{q,p}\right), 
\end{eqnarray}
and defining $\pi =\pi^{ij}\eta_{ij}$ one gets,
\begin{equation}
  \label{eq:18}
  2\left(\pi_{ij} -\frac{1}{2} \pi \eta_{ij}\right) = h_{ij,0} -
  N_{i,j}-N_{j,i}. 
\end{equation}
In terms of these the quantities the second order Lagrangian takes the
form,
\begin{equation}
  \label{eq:19}
  {\cal L} =\pi^{ij} h_{ij,0} - \left(1-\frac{h_{00}}{2}\right) R^0 - N_i R^i,
\end{equation}
where,
\begin{eqnarray}
  \label{eq:20}
  R^0&=&-\sqrt{1+h_a{}^a}\, {}^3\!R + \left(\pi^{ij}\pi_{ij} -\frac{1}{2}
         \pi^2\right),\\
R^i &=& -2 \pi^{ij}_{,j}.
\end{eqnarray}

As the action is partially gauge fixed, the total Hamiltonian
includes a true Hamiltonian plus a linear combination of constraints,
leading to a Lagrangian,
\begin{equation}
  \label{eq:21}
  {\cal L}_2 = \pi^{ij} \partial_0 h_{ij} -\left(\pi^{ij}\pi_{ij}
    -\frac{1}{2} \pi^2\right) -\frac{h^{00}}{2} {}^3 R + 2 N_i
  \pi^{ij}_{,j},
\end{equation}
which allows to determine the Hamiltonian and the constraints and to
define Poisson brackets that lead to canonical equations for $h_{ij}$
and $\pi^{ij}$,
\begin{equation}
  \label{eq:22}
  \left\{ h_{ij}(x),\pi^{kl}(y)\right\} = \delta_{ij}{}^{kl} \delta^3(x-y),
\end{equation}
with $\delta_{ij}{}^{kl} = \delta_i^k \delta_j^l+\delta_i^l\delta_j^k$.
Therefore,
\begin{eqnarray}
  \label{eq:23}
\left\{ h_{ij}(x), h^{kl}_{,0}(y)\right\}&=& 2\kappa \left\{
  h_{ij}(x),\pi^{kl}(y)-\frac{1}{2} \pi(y) \eta^{kl}\right\}
                                             \nonumber\\ &=& \kappa \left(2
  \delta_{ij}{}^{kl} - \delta_{ij}{}^{mn} \eta_{mn} \eta^{kl}\right)
                                             \delta^3(x-y)\nonumber\\
&=&\kappa\left(2 \delta_i^k \delta_j^l +2 \delta_i^l \delta_j^k-2\eta_{ij}\eta^{kl}\right)\delta^3(x-y),\label{fundpoisson}
\end{eqnarray}
with $x,y$ Riemann or Fermi normal coordinates.

To compute the Poisson brackets between intrinsic Riemann tensors,
we use Riemann normal coordinates $x^a$ around a point $p_0$. 
Given an intrinsic path
$y^\alpha(s)$ the corresponding curve in Riemann normal coordinates is
$\gamma^a(y,R)$. We are interested in computing the Poisson bracket
between the components of Riemann tensors in $p_0$ and $p$.

\begin{figure}[h]
\includegraphics[height=5cm]{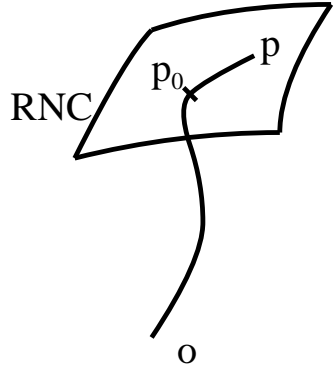}
\caption{The Riemann normal coordinates used in the computations of
  the Poisson brackets.}
\label{fig1}
\end{figure}

So we have an intrinsic path and we know the Riemann tensor at $p_0$.
The metric there is flat and the connection vanishes. In a
neighborhood of that point we work in Riemann normal coordinates, and
we wish to relate the Riemann tensor in such coordinates at $p$ with
the intrinsic Riemann tensor there. 

We have the curves $y^\alpha = s u^\alpha$ with $u^\alpha$ a unitary
vector at $p_0$ and $s$ the geodesic distance. In Riemann normal
coordinates they have the same expression, $x^a =s u^a$ (we are using
Latin letters for the Riemann coordinates and Greek letters for the
intrinsic ones). The
metric at $p_0$ is flat and the tetrads are
$e_\alpha^a=\delta_\alpha^a$, and $e^\alpha_a =\delta^\alpha_a$. The
tetrads at $p$ are given by the parallel transport (see \cite{Nesterov}),
\begin{eqnarray}
  \label{eq:37}
  e^{(\alpha)}_a\left(p\right) &=& \delta^\alpha_a + \frac{1}{s}
                                   \int_0^{s_p} d\sigma \int_0^\sigma
                                   d\sigma' \sigma' e^{(\alpha)}_b
                                   R^b{}_{cda} u^c u^d\nonumber\\
  &=&\delta^{(\alpha)}_a +\frac{1}{s} \int _0^{s_p} d\sigma
      \int_0^\sigma d\sigma' \sigma' u^c u^d + O\left(R^2\right)=O\left(s_p^4\right),
\end{eqnarray}
and
\begin{equation}
  \label{eq:38}
  R_{\alpha\beta\gamma\delta}\left(\pi_o^{p_o} \circ
    \pi_{p_o}^p\right) = e_\alpha^a e_\beta^b e_\gamma^c e_\delta^d
  R_{abcd}\left(x=u s_p\right),
\end{equation}
where the tetrads are evaluated at $\left(\pi_o^{p_o} \circ
    \pi_{p_o}^p\right) $. Defining the linearized tetrad at that
  point, $e_\alpha^a=\delta_\alpha^a +\delta e_\alpha^a$, we wish to
  compute the Poisson bracket between the Riemann tensor at $p$ and
  $p_0$ in intrinsic coordinates,
\begin{equation}
  \label{eq:39}
\left\{R_{\mu\nu\lambda\rho}(\pi_o^{p_0}),R_{\alpha\beta\gamma\delta}(\pi_0^{p_0}\circ \pi_{p_0}^p) \right\}.
\end{equation}
Notice that we are restricting ourselves to paths going to $p$ and
$p_0$ that are continuations of each other. One could have reached $p$
and $p_0$ by different paths. Such a calculation can be inferred from
the present one by adding additional loops to the paths. However, all
the relevant information for the general computation is present in the
one shown in equation (\ref{eq:39}) so we will concentrate on it.

Given that at $p_0$ the intrinsic and Riemann normal coordinates
coincide, we have,
\begin{equation}
  \label{eq:44}
  R_{\mu\nu\lambda\rho}(\pi_o^{p_0})=\delta_\mu^m \delta_\nu^n
  \delta_\lambda^l \delta_\rho^r R_{mnlr}(p_0).
\end{equation}
Also, given that and the expansion of the tetrads, we get,
\begin{equation}
  \label{eq:45}
  R_{\alpha\beta\gamma\delta} (\pi_o^{p_0}\circ \pi_{p_0}^p)=
  \delta_\alpha^a \delta_\beta^b \delta_\gamma^c \delta_\delta^d
R_{abcd} (p)+ e_\alpha^a \delta_\beta^b \delta_\gamma^c
\delta_\delta^d R_{abcd}(p) +\ldots
\end{equation}
where the dots mean the repetition of the same construction for the other
indices

With this we can reduce the computation of the Poisson bracket of the
Riemann tensor in intrinsic coordinates to that in Riemann normal
coordinates using the Poisson brackets we already presented. We will
discuss this in the following section.

\section{Explicit computation}

\subsection{General Poisson bracket to be computed}

We would like to compute two fundamental non-trivial Poisson brackets. The first
one, which we call $P_1$,  involves the Riemann tensor with one zeroth index and the Riemann
tensor with spatial indices. The second one, which we denote by $P_2$,
involves
two Riemann tensors with one zeroth components. The other Poisson
brackets can be readily derived from these ones using the equations of motion
as they involve second time derivatives. As is usual in Poisson
bracket computations, it is convenient to smear the functions, at
least for one of the terms, we do so with a test function $\phi(y)$, 
\begin{align}
P_{1}= & \left\{ R_{0IJK}\left(x\right),\int
         d^{3}y\,\phi\left(y\right)R_{ABCD}\left(y\right)\right.\nonumber\\
&+\int d^{3}y\,\phi\left(y\right)\delta e_{A}{}^{m}\left(y\right)R_{mBCD}\left(y\right)+\int d^{3}y\,\phi\left(y\right)\delta e_{B}{}^{m}\left(y\right)R_{AmCD}\left(y\right)\nonumber \\
 & \left.+\int d^{3}y\,\phi\left(y\right)\delta e_{C}{}^{m}\left(y\right)R_{ABmD}\left(y\right)+\int d^{3}y\,\phi\left(y\right)\delta e_{D}{}^{m}\left(y\right)R_{ABCm}\left(y\right)\right\} ,\label{Eq:Full-Poisson-1}
\end{align}
and
\begin{align}
P_{2}= & \left\{ R_{0IJK}\left(x\right),\int
         d^{3}y\,\phi\left(y\right)R_{0BCD}\left(y\right)\right.\nonumber\\
&+\int d^{3}y\,\phi\left(y\right)\delta e_{0}{}^{m}\left(y\right)R_{mBCD}\left(y\right)+\int d^{3}y\,\phi\left(y\right)\delta e_{B}{}^{m}\left(y\right)R_{0mCD}\left(y\right)\nonumber \\
 & \left.+\int d^{3}y\,\phi\left(y\right)\delta e_{C}{}^{m}\left(y\right)R_{0BmD}\left(y\right)+\int d^{3}y\,\phi\left(y\right)\delta e_{D}{}^{m}\left(y\right)R_{0BCm}\left(y\right)\right\} ,\label{Eq:Full-Poisson-2}
\end{align}
where $A,B,C,\ldots$ are spatial indices while $a,b,c,\ldots$ are
spacetime indices, both in Riemann normal coordinates. These two brackets
can be considered as special cases of
\begin{align}
P= & \left\{ R_{0IJK}\left(x\right),\int
     d^{3}y\,\phi\left(y\right)R_{aBCD}\left(y\right)\right.\nonumber\\
&+\int d^{3}y\,\phi\left(y\right)\delta e_{a}{}^{m}\left(y\right)R_{mBCD}\left(y\right)+\int d^{3}y\,\phi\left(y\right)\delta e_{B}{}^{m}\left(y\right)R_{amCD}\left(y\right)\nonumber \\
 & \left.+\int d^{3}y\,\phi\left(y\right)\delta e_{C}{}^{m}\left(y\right)R_{aBmD}\left(y\right)+\int d^{3}y\,\phi\left(y\right)\delta e_{D}{}^{m}\left(y\right)R_{aBCm}\left(y\right)\right\} .\label{Eq:Full-Poisson-gen}
\end{align}

Using the notations introduced by Mandelstam,
\begin{align}
\underset{i\leftrightarrow j}{A}\,f_{ij}= & f_{ij}-f_{ji},\label{eq:mand-not-1}\\
\underset{a\leftrightarrow b}{S}\,F_{ab}= & F_{ab}+F_{ba}-\eta_{ab}F_{r}{}^{r},\label{eq:mand-not-2}
\end{align}
it can be written as
\begin{align}
P= & \int d^{3}y\,\phi\left(y\right)\left\{ R_{0IJK}\left(x\right),R_{aBCD}\left(y\right)\right\} \nonumber \\
 & -\underset{a\leftrightarrow B}{A}\int d^{3}y\,\phi\left(y\right)\left\{ R_{0IJK}\left(x\right),\delta e_{B}{}^{m}\left(y\right)R_{maCD}\left(y\right)\right\} \nonumber \\
 & -\underset{C\leftrightarrow D}{A}\int d^{3}y\,\phi\left(y\right)\left\{ R_{0IJK}\left(x\right),\delta e_{D}{}^{m}\left(y\right)R_{mCaB}\left(y\right)\right\} ,\label{Eq:Full-Poisson-gen-sym-Mandl}
\end{align}
or, expanding the products,
\begin{align}
P= & \int d^{3}y\,\phi\left(y\right)\left\{ R_{0IJK}\left(x\right),R_{aBCD}\left(y\right)\right\} \nonumber \\
 & -\underset{a\leftrightarrow B}{A}\int d^{3}y\,\phi\left(y\right)\left\{ R_{0IJK}\left(x\right),\delta e_{B}{}^{m}\left(y\right)\right\} R_{maCD}\left(y\right)\nonumber \\
 & -\underset{a\leftrightarrow B}{A}\int d^{3}y\,\phi\left(y\right)\delta e_{B}{}^{m}\left(y\right)\left\{ R_{0IJK}\left(x\right),R_{maCD}\left(y\right)\right\} \nonumber \\
 & -\underset{C\leftrightarrow D}{A}\int d^{3}y\,\phi\left(y\right)\left\{ R_{0IJK}\left(x\right),\delta e_{D}{}^{m}\left(y\right)\right\} R_{mCaB}\left(y\right)\nonumber \\
 & -\underset{C\leftrightarrow D}{A}\int d^{3}y\,\phi\left(y\right)\delta e_{D}{}^{m}\left(y\right)\left\{ R_{0IJK}\left(x\right),R_{mCaB}\left(y\right)\right\} .\label{Eq:Full-Psn-expnd}
\end{align}
Hence, in principle we need to compute the following Poisson brackets,
\begin{align}
\Theta_{1}= & \left\{ R_{0IJK}\left(x\right),R_{ABCD}\left(y\right)\right\} ,\label{eq:Theta-1}\\
\Theta_{2}= & \left\{ R_{0IJK}\left(x\right),R_{0BCD}\left(y\right)\right\} ,\label{eq:Theta-2}\\
\Theta_{3}= & \left\{ R_{0IJK}\left(x\right),R_{0B0D}\left(y\right)\right\} ,\label{eq:Theta-3}\\
\Theta_{4}= & \left\{ R_{0IJK}\left(x\right),\delta e_{B}{}^{M}\left(y\right)\right\} ,\label{eq:Theta-4}\\
\Theta_{5}= & \left\{ R_{0IJK}\left(x\right),\delta e_{B}{}^{0}\left(y\right)\right\} ,\label{eq:Theta-5}\\
\Theta_{6}= & \left\{ R_{0IJK}\left(x\right),\delta e_{0}{}^{0}\left(y\right)\right\} ,\label{eq:Theta-6}\\
\Theta_{7}= & \left\{ R_{0IJK}\left(x\right),\delta e_{0}{}^{M}\left(y\right)\right\} \label{eq:Theta-7}
\end{align}
to get to the final results. Now we compute the terms (\ref{eq:Theta-1})
to (\ref{eq:Theta-7}) and then substitute them into
(\ref{Eq:Full-Psn-expnd}) to
find the Poisson brackets. 

\subsection{Relation between partial and Mandelstam derivatives}

\begin{figure}[t]
\vspace{-4cm}
\includegraphics[height=20cm]{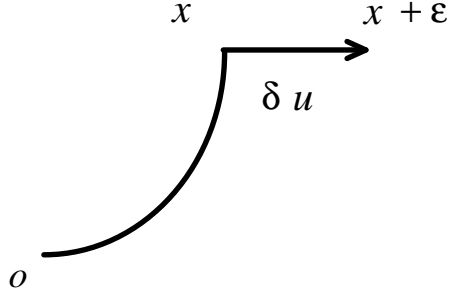}
\vspace{-12cm}
\caption{The path that defines the Mandelstam derivative,
  $\pi_E=\pi_o^x\circ \delta u$}
\label{mandelstamder}
\end{figure}

\begin{figure}[t]
\includegraphics[height=6cm]{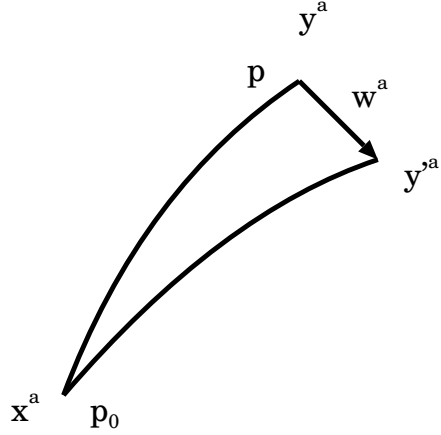}
\caption{The quantities involved in the Mandelstam derivative applied
  to our case.}
\label{mandelstamderint}
\end{figure}

The expressions we have involve partial derivatives, 
which correspond to Riemann normal coordinates in the calculations we are interested in.
We need to translate those into path dependent
derivatives like the Mandelstam derivatives we discussed in our
previous paper. 
Here we briefly recall the definition of such derivatives (for more details
see for instance \cite{gapubook}). Given a path
dependent function $\Psi(\pi_o^x)$ and a vector $u^\alpha$, the Mandelstam derivative is
obtained by considering its change when the path is extended from $x$
to $x+\epsilon u$ through an infinitesimal path $\delta u$ shown in
figure (\ref{mandelstamder}),
\begin{equation}
\Psi(\pi_o^x \circ \delta u) = (1 + \epsilon u^\alpha D_\alpha) \Psi(\pi_o^x).
\end{equation}
In our case we need to adapt this definition to the situation we wish
to consider, illustrated in figure (\ref{mandelstamderint}). Notice
that although it appears similar to (\ref{mandelstamder}) we need to
take into account that $w^\alpha$ is rotated since the Mandelstam
derivative must be referred to the frame parallel transported along a
geodesic from $x$ to $y$ with $u^\alpha=y^\alpha-x^\alpha$.

This yields,
\begin{equation}
\frac{\partial}{\partial y^{a}}=D_{a}^{y}-\frac{1}{6}R_{bac}{}^{n}\left(y^{b}-x^{b}\right)\left(y^{c}-x^{c}\right)D_{n}^{y},\label{eq:1part-der-mand-der}
\end{equation}
where by $D_a^y$ we denote the Mandelstam derivative acting at point
$y$. 

Hence for the time derivatives we have
\begin{align}
\frac{\partial}{\partial y^{0}}= & D_{0}^{y}-\frac{1}{6}R_{b0c}{}^{n}\left(y^{b}-x^{b}\right)\left(y^{c}-x^{c}\right)D_{n}^{y},\label{eq:Mandl-partial-0-y}\\
\frac{\partial}{\partial x^{0}}= & D_{0}^{x}-\frac{1}{6}R_{b0c}{}^{n}\left(x^{b}-x^{b}\right)\left(x^{c}-x^{c}\right)D_{n}^{x}=D_{0}^{x}.\label{eq:Mandl-partial-0-x}
\end{align}
and we also get a relationship we will need in future computations:
\begin{equation}
\frac{\partial}{\partial x^{J}}\frac{\partial}{\partial y^{c}}\frac{\partial}{\partial y^{a}}=D_{J}^{x}D_{c}^{y}D_{a}^{y}+\frac{1}{6}R_{Jac}{}^{n}D_{n}^{y}.\label{eq:Mandl-3deriv}
\end{equation}

\subsection{Terms $\Theta_{1}$ - $\Theta_{3}$}

The terms (\ref{eq:Theta-1})-(\ref{eq:Theta-3}) become
\begin{align}
\Theta_{1}= & \underset{J\leftrightarrow K}{A}\underset{C\leftrightarrow D}{A}\left\{ \frac{\partial}{\partial x^{J}}\frac{\partial}{\partial x^{[I}}h_{0]K}\left(x\right),\frac{\partial}{\partial y^{C}}\frac{\partial}{\partial y^{[B}}h_{A]D}\left(y\right)\right\} ,\label{eq:Theta-1-expr-2}\\
\Theta_{2}= & \underset{J\leftrightarrow K}{A}\underset{C\leftrightarrow D}{A}\left\{ \frac{\partial}{\partial x^{J}}\frac{\partial}{\partial x^{[I}}h_{0]K}\left(x\right),\frac{\partial}{\partial y^{C}}\frac{\partial}{\partial y^{[B}}h_{0]D}\left(y\right)\right\} ,\label{eq:Theta-2-expr-2}\\
\Theta_{3}= & \underset{J\leftrightarrow K}{A}\underset{0^{\prime}\leftrightarrow D}{A}\left\{ \frac{\partial}{\partial x^{J}}\frac{\partial}{\partial x^{[I}}h_{0]K}\left(x\right),\frac{\partial}{\partial y^{0^{\prime}}}\frac{\partial}{\partial y^{[B}}h_{0]D}\left(y\right)\right\} ,\label{eq:Theta-3-expr-2}
\end{align}
where in the last term we distinguish the $0^{\prime}$ index that
is being interchanged with $D$ from other $0$ indices.

The Poisson bracket in term (\ref{eq:Theta-1-expr-2}) can be written
as
\begin{align}
\Theta_1=\left\{ \frac{\partial}{\partial
  x^{J}}\frac{\partial}{\partial
  x^{[I}}h_{0]K}\left(x\right),\frac{\partial}{\partial
  y^{C}}\frac{\partial}{\partial y^{[B}}h_{A]D}\left(y\right)\right\}.
\end{align} 
Using the Poisson bracket in Riemann normal coordinates
(\ref{fundpoisson}) one can easily obtain,
\begin{align}
\left\{ \frac{\partial}{\partial x^{0}}h_{IK}\left(x\right),h_{bd}\left(y\right)\right\} = & -2\kappa\left(\eta_{bI}\eta_{dK}+\eta_{bK}\eta_{dI}-\eta_{bd}\eta_{IK}\right)\delta^{(3)}\left(y-x\right)\nonumber \\
= & -2\kappa\underset{I\leftrightarrow K}{S}\eta_{bK}\eta_{dI}\delta^{(3)}\left(y-x\right),\label{eq:PSN-princip-d0}
\end{align}
noticing that the lapse and shift commute with the canonical
variables.

With these expressions we have,
\begin{equation}
\Theta_{1}=-\frac{\kappa}{2}\underset{A\leftrightarrow B}{A}\underset{C\leftrightarrow D}{A}\underset{J\leftrightarrow K}{A}\underset{I\leftrightarrow K}{S}\frac{\partial}{\partial x^{J}}\frac{\partial}{\partial y^{C}}\frac{\partial}{\partial y^{A}}\left[\eta_{BI}\eta_{DK}\delta^{(3)}\left(y-x\right)\right].\label{eq:Theta-1-expr-3}
\end{equation}
However, in terms of Mandelstam derivatives, using (\ref{eq:Mandl-3deriv}),
we get
\begin{equation}
\frac{\partial}{\partial x^{J}}\frac{\partial}{\partial y^{c}}\frac{\partial}{\partial y^{a}}\delta^{(3)}\left(y-x\right)=D_{J}^{x}D_{c}^{y}D_{a}^{y}\delta^{(3)}\left(y-x\right)+\frac{1}{6}R_{Jac}{}^{n}D_{n}^{y}\delta^{(3)}\left(y-x\right).
\end{equation}
Using this in (\ref{eq:Theta-1-expr-3}) yields
\begin{align}
\Theta_{1}= & -\frac{\kappa}{2}\underset{A\leftrightarrow B}{A}\underset{C\leftrightarrow D}{A}\underset{J\leftrightarrow K}{A}\underset{I\leftrightarrow K}{S}D_{J}^{x}D_{C}^{y}D_{A}^{y}\left[\eta_{BI}\eta_{DK}\delta^{(3)}\left(y-x\right)\right]\nonumber \\
 & -\frac{\kappa}{12}\underset{A\leftrightarrow B}{A}\underset{C\leftrightarrow D}{A}\underset{J\leftrightarrow K}{A}\underset{I\leftrightarrow K}{S}R_{JAC}{}^{n}D_{n}^{y}\left[\eta_{BI}\eta_{DK}\delta^{(3)}\left(y-x\right)\right].\label{eq:Theta-1-expr-fin}
\end{align}
It should be noted that in this expression the arguments of the Dirac
delta are really the paths going from $o$ to $x$ and $y$ so that the
Mandelstam derivative can act on them. We keep the usual notation
$\delta^{(3)}(x-y)$ as a shorthand in this and future similar
expressions. 

The Poisson bracket in term (\ref{eq:Theta-2-expr-2}) can be written
as
\begin{align}\Theta_2=
\left\{ \frac{\partial}{\partial x^{J}}\frac{\partial}{\partial x^{[I}}h_{0]K}\left(x\right),\frac{\partial}{\partial y^{C}}\frac{\partial}{\partial y^{[B}}h_{0]D}\left(y\right)\right\} \label{eq:Theta-2-expr-fin}
\end{align}
and one can check that it vanishes.

For the $\Theta_3$ term, (\ref{eq:Theta-3-expr-2}), we have
\begin{align}
\Theta_{3}=  \underset{J\leftrightarrow K}{A}\underset{0^{\prime}\leftrightarrow D}{A}\left\{ \frac{\partial}{\partial x^{J}}\frac{\partial}{\partial x^{[I}}h_{0]K}\left(x\right),\frac{\partial}{\partial y^{0^{\prime}}}\frac{\partial}{\partial y^{[B}}h_{0]D}\left(y\right)\right\}.
\end{align}

So we get
\begin{align}
\Theta_{3}= & \frac{\kappa}{2}\underset{J\leftrightarrow K}{A}\underset{I\leftrightarrow K}{S}\eta^{CF}\frac{\partial}{\partial x^{J}}\frac{\partial}{\partial y^{F}}\frac{\partial}{\partial y^{B}}\eta_{CI}\eta_{DK}\delta^{(3)}\left(y-x\right)\nonumber \\
 & -\frac{\kappa}{2}\underset{J\leftrightarrow K}{A}\underset{I\leftrightarrow K}{S}\eta^{CF}\frac{\partial}{\partial x^{J}}\frac{\partial}{\partial y^{F}}\frac{\partial}{\partial y^{C}}\eta_{BI}\eta_{DK}\delta^{(3)}\left(y-x\right)\nonumber \\
 & +\frac{\kappa}{2}\underset{J\leftrightarrow K}{A}\underset{I\leftrightarrow K}{S}\eta^{CF}\frac{\partial}{\partial x^{J}}\frac{\partial}{\partial y^{F}}\frac{\partial}{\partial y^{D}}\eta_{BI}\eta_{CK}\delta^{(3)}\left(y-x\right)\nonumber \\
 & -\frac{\kappa}{2}\underset{J\leftrightarrow K}{A}\underset{I\leftrightarrow K}{S}\eta^{cd}\frac{\partial}{\partial x^{J}}\frac{\partial}{\partial y^{D}}\frac{\partial}{\partial y^{B}}\eta_{cI}\eta_{dK}\delta^{(3)}\left(y-x\right).
\end{align}
Using (\ref{eq:Mandl-3deriv}) in above, we get, in terms of paths, 
\begin{align}
\Theta_{3}= & \frac{\kappa}{2}\underset{J\leftrightarrow K}{A}\underset{I\leftrightarrow K}{S}\eta^{CF}D_{J}^{x}D_{F}^{y}D_{B}^{y}\left[\eta_{CI}\eta_{DK}\delta^{(3)}\left(y-x\right)\right]\nonumber \\
 & -\frac{\kappa}{2}\underset{J\leftrightarrow K}{A}\underset{I\leftrightarrow K}{S}\eta^{CF}D_{J}^{x}D_{F}^{y}D_{C}^{y}\left[\eta_{BI}\eta_{DK}\delta^{(3)}\left(y-x\right)\right]\nonumber \\
 & +\frac{\kappa}{2}\underset{J\leftrightarrow K}{A}\underset{I\leftrightarrow K}{S}\eta^{CF}D_{J}^{x}D_{F}^{y}D_{D}^{y}\left[\eta_{BI}\eta_{CK}\delta^{(3)}\left(y-x\right)\right]\nonumber \\
 & -\frac{\kappa}{2}\underset{J\leftrightarrow K}{A}\underset{I\leftrightarrow K}{S}\eta^{cd}D_{J}^{x}D_{D}^{y}D_{B}^{y}\left[\eta_{cI}\eta_{dK}\delta^{(3)}\left(y-x\right)\right]\nonumber \\
 & +\frac{\kappa}{12}\underset{J\leftrightarrow K}{A}\underset{I\leftrightarrow K}{S}\eta^{CF}R_{JBF}{}^{n}D_{n}^{y}\left[\eta_{CI}\eta_{DK}\delta^{(3)}\left(y-x\right)\right]\nonumber \\
 & -\frac{\kappa}{12}\underset{J\leftrightarrow K}{A}\underset{I\leftrightarrow K}{S}\eta^{CF}R_{JCF}{}^{n}D_{n}^{y}\left[\eta_{BI}\eta_{DK}\delta^{(3)}\left(y-x\right)\right]\nonumber \\
 & +\frac{\kappa}{12}\underset{J\leftrightarrow K}{A}\underset{I\leftrightarrow K}{S}\eta^{CF}R_{JDF}{}^{n}D_{n}^{y}\left[\eta_{BI}\eta_{CK}\delta^{(3)}\left(y-x\right)\right]\nonumber \\
 & -\frac{\kappa}{12}\underset{J\leftrightarrow K}{A}\underset{I\leftrightarrow K}{S}\eta^{cd}R_{JBD}{}^{n}D_{n}^{y}\left[\eta_{cI}\eta_{dK}\delta^{(3)}\left(y-x\right)\right].\label{eq:Theta-3-expr-fin}
\end{align}

This contains second derivatives of the metric and therefore requires
the equations of motion, which we assume are in vacuum. Hence, we get,
\begin{align}
\Theta_{3}= & \frac{\kappa}{2}\underset{J\leftrightarrow K}{A}\underset{I\leftrightarrow K}{S}\eta^{CF}D_{J}^{x}D_{F}^{y}D_{B}^{y}\left[\eta_{CI}\eta_{DK}\delta^{(3)}\left(y-x\right)\right]\nonumber \\
 & -\frac{\kappa}{2}\underset{J\leftrightarrow K}{A}\underset{I\leftrightarrow K}{S}\eta^{CF}D_{J}^{x}D_{F}^{y}D_{C}^{y}\left[\eta_{BI}\eta_{DK}\delta^{(3)}\left(y-x\right)\right]\nonumber \\
 & +\frac{\kappa}{2}\underset{J\leftrightarrow K}{A}\underset{I\leftrightarrow K}{S}\eta^{CF}D_{J}^{x}D_{F}^{y}D_{D}^{y}\left[\eta_{BI}\eta_{CK}\delta^{(3)}\left(y-x\right)\right]\nonumber \\
 & -\frac{\kappa}{2}\underset{J\leftrightarrow K}{A}\underset{I\leftrightarrow K}{S}\eta^{cd}D_{J}^{x}D_{D}^{y}D_{B}^{y}\left[\eta_{cI}\eta_{dK}\delta^{(3)}\left(y-x\right)\right]\nonumber \\
 & +\frac{\kappa}{12}\underset{J\leftrightarrow K}{A}\underset{I\leftrightarrow K}{S}\eta^{CF}R_{JBF}{}^{n}D_{n}^{y}\left[\eta_{CI}\eta_{DK}\delta^{(3)}\left(y-x\right)\right]\nonumber \\
 & -\frac{\kappa}{12}\underset{J\leftrightarrow K}{A}\underset{I\leftrightarrow K}{S}\eta^{CF}R_{JCF}{}^{n}D_{n}^{y}\left[\eta_{BI}\eta_{DK}\delta^{(3)}\left(y-x\right)\right]\nonumber \\
 & +\frac{\kappa}{12}\underset{J\leftrightarrow K}{A}\underset{I\leftrightarrow K}{S}\eta^{CF}R_{JDF}{}^{n}D_{n}^{y}\left[\eta_{BI}\eta_{CK}\delta^{(3)}\left(y-x\right)\right]\nonumber \\
 & -\frac{\kappa}{12}\underset{J\leftrightarrow K}{A}\underset{I\leftrightarrow K}{S}\eta^{cd}R_{JBD}{}^{n}D_{n}^{y}\left[\eta_{cI}\eta_{dK}\delta^{(3)}\left(y-x\right)\right].
\end{align}

\subsection{Terms $\Theta_{4}$ - $\Theta_{7}$}

In this case, we note that
\begin{equation}
\delta e_{a}{}^{m}=\frac{1}{s}\int_{0}^{s}d\sigma\int_{0}^{\sigma}d\sigma^{\prime}\,\sigma^{\prime}R^{m}{}_{cda}u^{c}u^{d}(y'),
\end{equation}
where $y'{}^a=y^a+\left(\sigma'-\sigma\right) u^a$ and
$u^a=\left(y^a-x^a\right)/\sigma$.

If we now expand $R^{m}{}_{cda}$ around $y$, with,
$y'{}^{l}=y^{l}+\left(\sigma'-\sigma\right)u^l=y^l+\epsilon^l$, we get,
\begin{equation}
R^{m}{}_{cda}\left(y'\right)=R^{m}{}_{cda}\left(y\right)+\partial_{h}R^{m}{}_{cda}(y)\epsilon^{h}+\mathcal{O}\left(\epsilon^{2}\right),
\end{equation}
which yields $\delta e_{a}{}^{m}$ as
\begin{align}
\delta e_{a}{}^{m}\approx & \frac{1}{s}\int_{0}^{s}d\sigma\int_{0}^{\sigma}d\sigma^{\prime}\,\sigma^{\prime}\left(R^{m}{}_{cda}\left(y\right)+\left.\partial_{h}R^{m}{}_{cda}\right|_{y}\epsilon^{h}+\left.\partial_{j}\partial_{h}R^{m}{}_{cda}\right|_{y}\epsilon^{h}\epsilon^{j}\right)u^{c}u^{d}\nonumber \\
= & R^{m}{}_{cda}\left(y\right)u^{c}u^{d}\left(\frac{1}{s}\int_{0}^{s}d\sigma\int_{0}^{\sigma}d\sigma^{\prime}\,\sigma^{\prime}\right)\nonumber \\
= & \frac{1}{6}R^{m}{}_{cda}\left(y\right)\left(y^{c}-x^{c}\right)\left(y^{d}-x^{d}\right)
\end{align}
where we neglect terms of higher order in epsilon and we have also used
\begin{equation}
\frac{1}{s^{3}}\int_{0}^{s}d\sigma\int_{0}^{\sigma}d\sigma^{\prime}\,\sigma^{\prime}=\frac{1}{s^{3}}\frac{1}{6}s^{3}=\frac{1}{6}.\label{eq:K12}
\end{equation}
Hence, we can write
\begin{align}
\delta e_{a}{}^{m}\approx & \frac{1}{6}R^{m}{}_{cda}\left(y\right)\left(y^{c}-x^{c}\right)\left(y^{d}-x^{d}\right)\nonumber \\
\approx & -\frac{1}{6}\eta^{mf}R_{DaCf}\left(y\right)\left(y^{C}-x^{C}\right)\left(y^{D}-x^{D}\right),\label{eq:delta-e-approx}
\end{align}
where we have assumed that the path going from $x$ to $y$ is spatial, so spatial indices need only be
considered. 
Thus the Poisson brackets in (\ref{eq:Theta-4}) - (\ref{eq:Theta-6})
can be written as
\begin{align}
\left\{ R_{0IJK}\left(x\right),\delta e_{a}{}^{m}\left(y\right)\right\} = & -\frac{1}{6}\eta^{mf}\left\{ R_{0IJK}\left(x\right),R_{DaCf}\left(y\right)\right\} \left(y^{C}-x^{C}\right)\left(y^{D}-x^{D}\right).\label{eq:thet-4-6-genr}
\end{align}
Now we compute $\Theta_{4}$ to $\Theta_{6}$ using this expression.

The $\Theta_{4}$ term corresponds to (\ref{eq:thet-4-6-genr}) with $m\to M,\,a\to B$
and $f\to F$, where the last one is a consequence of $\eta^{0I}=0$
for the outside $\eta^{mf}$. Hence we have
\begin{align}
\Theta_{4}= & \left\{ R_{0IJK}\left(x\right),\delta e_{B}{}^{M}\left(y\right)\right\} \nonumber \\
= & -\frac{1}{6}\eta^{MF}\left\{ R_{0IJK}\left(x\right),R_{DBCF}\left(y\right)\right\} \left(y^{C}-x^{C}\right)\left(y^{D}-x^{D}\right)\nonumber \\
= & -\frac{1}{6}\eta^{MF}\Theta_{1}\bigg|_{D\to F,A\to D}\left(y^{C}-x^{C}\right)\left(y^{D}-x^{D}\right)\nonumber \\
= & \eta^{MF}\frac{\kappa}{12}\left(\underset{D\leftrightarrow B}{A}\underset{C\leftrightarrow F}{A}\underset{J\leftrightarrow K}{A}\underset{I\leftrightarrow K}{S}D_{J}^{x}D_{C}^{y}D_{D}^{y}\left[\eta_{BI}\eta_{FK}\delta^{(3)}\left(y-x\right)\right]\right)\left(y^{C}-x^{C}\right)\left(y^{D}-x^{D}\right)\nonumber \\
 & +\eta^{MF}\frac{\kappa}{12\cdot6}\left(\underset{D\leftrightarrow B}{A}\underset{C\leftrightarrow F}{A}\underset{J\leftrightarrow K}{A}\underset{I\leftrightarrow K}{S}R_{JDC}{}^{n}D_{n}^{y}\left[\eta_{BI}\eta_{FK}\delta^{(3)}\left(y-x\right)\right]\right)\left(y^{C}-x^{C}\right)\left(y^{D}-x^{D}\right),
\end{align}
where we have used the expression (\ref{eq:Theta-1}). The notation in
$\Theta_1$ means substituting $D$ for $F$ and $A$ for $D$ in it. We
will use a similar notation in the following terms.

The $\Theta_{5}$ term corresponds to (\ref{eq:thet-4-6-genr}) with $m\to0,\,a\to B$,
and as a consequence $f\to0$. Hence
\begin{align}
\Theta_{5}= & \left\{ R_{0IJK}\left(x\right),\delta e_{B}{}^{0}\left(y\right)\right\} \nonumber \\
= & -\frac{1}{6}\eta^{00}\left\{ R_{0IJK}\left(x\right),R_{DBC0}\left(y\right)\right\} \left(y^{C}-x^{C}\right)\left(y^{D}-x^{D}\right)\nonumber \\
= & \frac{1}{6}\left\{ R_{0IJK}\left(x\right),R_{0CBD}\left(y\right)\right\} \left(y^{C}-x^{C}\right)\left(y^{D}-x^{D}\right)\nonumber \\
= & \frac{1}{6}\Theta_{2}\bigg|_{B\leftrightarrow C}\left(y^{C}-x^{C}\right)\left(y^{D}-x^{D}\right)\nonumber \\
= & 0\label{eq:Theta-5-fin}
\end{align}

The $\Theta_{6}$ term corresponds to (\ref{eq:thet-4-6-genr}) with $m\to0,\,a\to0$,
and as a consequence $f\to0$. Hence using \ref{eq:delta-e-approx}
we get
\begin{align}
\Theta_{6}= & \left\{ R_{0IJK}\left(x\right),\delta e_{0}{}^{0}\left(y\right)\right\} \nonumber \\
= & \left\{ R_{0IJK}\left(x\right),\frac{1}{6}R_{D0C0}\left(y\right)\left(y^{C}-x^{C}\right)\left(y^{D}-x^{D}\right)\right\} \nonumber \\
= & \frac{1}{6}\left(y^{C}-x^{C}\right)\left(y^{D}-x^{D}\right)\Theta_{3}\bigg|_{B\to C}.
\end{align}

The $\Theta_{7}$ term corresponds to (\ref{eq:thet-4-6-genr}) with $m\to M,\,a\to0$,
and as a consequence and $f\to F$. Hence using \ref{eq:delta-e-approx}
we get

\begin{align*}
\Theta_{7}= & \left\{ R_{0IJK}\left(x\right),\delta e_{0}{}^{M}\left(y\right)\right\} \\
= & -\frac{1}{6}\eta^{MF}\left(y^{C}-x^{C}\right)\left(y^{D}-x^{D}\right)\left\{ R_{0IJK}\left(x\right),R_{D0CF}\left(x\right)\right\} \\
= & \frac{1}{6}\eta^{MF}\left(y^{C}-x^{C}\right)\left(y^{D}-x^{D}\right)\Theta_{2}\bigg|_{D\to F,B\to D}\\
= & 0
\end{align*}

\section{Full Poisson brackets}

Now that we have all the necessary terms, we put them together to
get the first Poisson bracket (\ref{Eq:Full-Poisson-1}). Using
(\ref{Eq:Full-Psn-expnd}), and some of symmetries of
$A$ and $S$ operators and the symmetries of the Riemann tensor,
we get for (\ref{Eq:Full-Poisson-1}),
\begin{align}
P_{1}=P\big|_{a=A}= & \int d^{3}y\,\phi\left(y\right)\Theta_{1}\nonumber \\
 & -\underset{A\leftrightarrow B}{A}\int d^{3}y\,\phi\left(y\right)\Theta_{4}R_{MACD}\left(y\right)\nonumber \\
 & +\underset{A\leftrightarrow B}{A}\int d^{3}y\,\phi\left(y\right)\delta e_{B}{}^{M}\left(y\right)\Theta_{1}\bigg|_{B\to M}\nonumber \\
 & -\underset{C\leftrightarrow D}{A}\int d^{3}y\,\phi\left(y\right)\Theta_{4}\bigg|_{B\to D}R_{MCAB}\left(y\right)\nonumber \\
 & +\underset{C\leftrightarrow D}{A}\int d^{3}y\,\phi\left(y\right)\delta e_{D}{}^{M}\left(y\right)\Theta_{1}\bigg|_{D\to M}.
\end{align}

Recalling (\ref{eq:45}), we obtain the first Poisson bracket
\begin{align}
P_{1}=& \left\{ R_{0IJK}\left(\pi_o^x\right),R_{ABCD}\left(\pi_o^x\circ  \pi_x^y\right)\right\}\nonumber\\
=& -\frac{\kappa}{2}\underset{A\leftrightarrow B}{A}\underset{C\leftrightarrow D}{A}\underset{J\leftrightarrow K}{A}\underset{I\leftrightarrow K}{S}D_{J}^{x}D_{C}^{x}D_{A}^{x}\left[\delta^3\left(x-y\right)\right]\eta_{BI}\eta_{DK}\nonumber \\
 & +\kappa\underset{A\leftrightarrow B}{A}\underset{J\leftrightarrow K}{A}\underset{I\leftrightarrow K}{S}D_{J}^{x}\left[\delta^3\left(x-y\right)R^{F}{}_{ADC}\left(\pi_o^x\right)\right]\eta_{BI}\eta_{FK}\nonumber \\
 & +\frac{\kappa}{4}\underset{A\leftrightarrow B}{A}\underset{J\leftrightarrow K}{A}\underset{I\leftrightarrow K}{S}D_{J}^{x}\left[\delta^3\left(x-y\right)R^{F}{}_{CDA}\left(\pi_o^x\right)\right]\eta_{BI}\eta_{FK}\nonumber \\
 & -\frac{\kappa}{4}\underset{C\leftrightarrow D}{A}\underset{J\leftrightarrow K}{A}\underset{I\leftrightarrow K}{S}D_{J}^{x}\left[\delta^3\left(x-y\right)R^{F}{}_{CBA}\left(\pi_o^x\right)\right]\eta_{FI}\eta_{DK}\nonumber \\
 & +\kappa\underset{C\leftrightarrow D}{A}\underset{J\leftrightarrow K}{A}\underset{I\leftrightarrow K}{S}D_{J}^{x}\left[\delta^3\left(x-y\right)R^{F}{}_{CBA}\left(\pi_o^x\right)\right]\eta_{FI}\eta_{DK}\nonumber \\
 & -\frac{\kappa}{4}\underset{A\leftrightarrow B}{A}\underset{J\leftrightarrow K}{A}\underset{I\leftrightarrow K}{S}D_{F}^{x}\left[\delta^3\left(x-y\right)R^{F}{}_{ADC}\left(\pi_o^x\right)\right]\eta_{BI}\eta_{JK}\nonumber \\
 & -\frac{\kappa}{4}\underset{C\leftrightarrow D}{A}\underset{J\leftrightarrow K}{A}\underset{I\leftrightarrow K}{S}D_{F}^{x}\left[\delta^3\left(x-y\right)R^{F}{}_{CBA}\left(\pi_o^x\right)\right]\eta_{DI}\eta_{JK}\nonumber \\
 & +\frac{\kappa}{12}\underset{A\leftrightarrow B}{A}\underset{C\leftrightarrow D}{A}\underset{J\leftrightarrow K}{A}\underset{I\leftrightarrow K}{S}D_{F}^{x}\left[\delta^3\left(x-y\right)R^{F}{}_{CAJ}\left(\pi_o^x\right)\right]\eta_{BI}\eta_{DK}\nonumber \\
 & +\frac{\kappa}{12}\underset{A\leftrightarrow B}{A}\underset{C\leftrightarrow D}{A}\underset{J\leftrightarrow K}{A}\underset{I\leftrightarrow K}{S}D_{F}^{x}\left[\delta^3\left(x-y\right)\left(R^{F}{}_{JBC}\left(\pi_o^x\right)+R^{F}{}_{CBJ}\left(\pi_o^x\right)\right)\right]\eta_{AI}\eta_{DK}\nonumber \\
 & -\frac{\kappa}{12}\underset{A\leftrightarrow B}{A}\underset{C\leftrightarrow D}{A}\underset{J\leftrightarrow K}{A}\underset{I\leftrightarrow K}{S}D_{F}^{x}\left[\delta^3\left(x-y\right)\left(R^{F}{}_{JDA}\left(\pi_o^x\right)+R^{F}{}_{ADJ}\left(\pi_o^x\right)\right)\right]\eta_{BI}\eta_{CK}\nonumber \\
 & -\frac{\kappa}{12}\underset{A\leftrightarrow B}{A}\underset{C\leftrightarrow D}{A}\underset{J\leftrightarrow K}{A}\underset{I\leftrightarrow K}{S}D_{A}^{x}\left[\delta^3\left(x-y\right)\left(R^{F}{}_{JBC}\left(\pi_o^x\right)+R^{F}{}_{CBJ}\left(\pi_o^x\right)\right)\right]\eta_{FI}\eta_{DK}\nonumber \\
 & +\frac{\kappa}{4}\underset{A\leftrightarrow B}{A}\underset{J\leftrightarrow K}{A}\underset{I\leftrightarrow K}{S}D_{A}^{x}\left[\delta^3\left(x-y\right)R^{F}{}_{JDC}\left(\pi_o^x\right)\right]\eta_{BI}\eta_{FK}\nonumber \\
 & -\frac{\kappa}{4}\underset{A\leftrightarrow B}{A}\underset{J\leftrightarrow K}{A}\underset{I\leftrightarrow K}{S}D_{B}^{x}\left[\delta^3\left(x-y\right)R^{F}{}_{ADC}\left(\pi_o^x\right)\right]\eta_{JI}\eta_{FK}\nonumber \\
 & +\frac{\kappa}{12}\underset{A\leftrightarrow B}{A}\underset{C\leftrightarrow D}{A}\underset{J\leftrightarrow K}{A}\underset{I\leftrightarrow K}{S}D_{C}^{x}\left[\delta^3\left(x-y\right)\left(R^{F}{}_{JDA}\left(\pi_o^x\right)+R^{F}{}_{ADJ}\left(\pi_o^x\right)\right)\right]\eta_{BI}\eta_{FK}\nonumber \\
 & -\frac{\kappa}{4}\underset{C\leftrightarrow D}{A}\underset{J\leftrightarrow K}{A}\underset{I\leftrightarrow K}{S}D_{C}^{x}\left[\delta^3\left(x-y\right)R^{F}{}_{JBA}\left(\pi_o^x\right)\right]\eta_{FI}\eta_{DK}\nonumber \\
 & -\frac{\kappa}{4}\underset{C\leftrightarrow D}{A}\underset{J\leftrightarrow K}{A}\underset{I\leftrightarrow K}{S}D_{D}^{x}\left[\delta^3\left(x-y\right)R_{ABC}{}^{F}\left(\pi_o^x\right)\right]\eta_{JI}\eta_{FK},\nonumber \\
\end{align}
where we did not include contributions due to the Ricci tensor as we
are assuming we are in vacuum. One can check that the above expression
for the Poisson bracket is compatible with the vacuum equations by
contracting it with $\eta^{IJ}$ and seeing that it vanishes. This is
most easily seen considering each of the terms in the expansion of
(4.3) and showing that their contraction with $\eta^{IJ}$ vanishes. Therefore,
\begin{equation}
  \label{eq:46}
  \left\{ R_{0K}\left(\pi_o^x\right),R_{ABCD}\left(\pi_o^x\circ  \pi_x^y\right)\right\}=0.
\end{equation}

Introducing the symmetrizer notation for  pairs of indices,
\begin{equation}
\underset{X\leftrightarrow Y}{T}f_{XY}=f_{XY}+f_{YX},
\end{equation}
the expression can be made more compact,
\begin{align}
P_{1}= & -\frac{\kappa}{2}\underset{A\leftrightarrow B}{A}\underset{C\leftrightarrow D}{A}\underset{J\leftrightarrow K}{A}\underset{I\leftrightarrow K}{S}D_{J}^{x}D_{C}^{x}D_{A}^{x}\left[\delta^3\left(x-y\right)\right]\eta_{BI}\eta_{DK}\nonumber \\
 & +\frac{\kappa}{12}\underset{A\leftrightarrow B}{A}\underset{C\leftrightarrow D}{A}\underset{J\leftrightarrow K}{A}\underset{I\leftrightarrow K}{S}D_{F}^{x}\left[\delta^3\left(x-y\right)R_{JAC}{}^{F}\left(\pi_o^x\right)\right]\eta_{BI}\eta_{DK}\nonumber \\
 & +\underset{A\leftrightarrow C;B\leftrightarrow D}{T}\left\{ \underset{A\leftrightarrow B}{A}\underset{J\leftrightarrow K}{A}\left[\kappa\underset{I\leftrightarrow K}{S}D_{J}^{x}\left[\delta^3\left(x-y\right)R^{F}{}_{ADC}\left(\pi_o^x\right)\right]\eta_{BI}\eta_{FK}\right.\right.\nonumber \\
 & -\frac{\kappa}{4}\underset{I\leftrightarrow K}{S}\underset{F\leftrightarrow B}{T}D_{F}^{x}\left[\delta^3\left(x-y\right)R^{F}{}_{ADC}\left(\pi_o^x\right)\right]\eta_{BI}\eta_{JK}\nonumber \\
 & +\frac{\kappa}{12}\underset{C\leftrightarrow D}{A}\underset{I\leftrightarrow K}{S}\underset{J\leftrightarrow C}{T}\underset{A\leftrightarrow C}{T}D_{J}^{x}\left[\delta^3\left(x-y\right)R^{F}{}_{ABC}\left(\pi_o^x\right)\right]\eta_{DI}\eta_{FK}\nonumber \\
 & \left.\left.+\frac{\kappa}{12}\underset{C\leftrightarrow D}{A}\underset{I\leftrightarrow K}{S}\underset{J\leftrightarrow C}{T}\underset{A\leftrightarrow F}{A}D_{A}^{x}\left[\delta^3\left(x-y\right)R^{F}{}_{JBC}\left(\pi_o^x\right)\right]\eta_{DI}\eta_{FK}\right]\right\}.\label{eq:full-final-PB-1}
\end{align}

We can now proceed to compute the second Poisson bracket. We are interested in computing, 
\begin{equation}
  \label{eq:4}
P_2=  \left\{ R_{0IJK}\left(\pi_o^x\right),R_{0BCD}\left(\pi_o^x\circ  \pi_x^y\right)\right\}.
\end{equation}
we have,
\begin{align}
P_{2}= & P\big|_{a=0}\nonumber \\
= & \int d^{3}y\,\phi\left(y\right)R_{0BCD}\left(y\right)\Theta_{6}\nonumber \\
 & -\int d^{3}y\,\phi\left(y\right)R_{M0CD}\left(y\right)\Theta_{4}\nonumber \\
 & +\int d^{3}y\,\phi\left(y\right)\delta e_{0}{}^{M}\left(y\right)\Theta_{1}\bigg|_{A\to M}\nonumber \\
 & -\underset{C\leftrightarrow D}{A}\int d^{3}y\,\phi\left(y\right)R_{MC0B}\left(y\right)\Theta_{4}\bigg|_{B\to D}\nonumber \\
 & -\underset{C\leftrightarrow D}{A}\int d^{3}y\,\phi\left(y\right)\delta e_{D}{}^{0}\left(y\right)\Theta_{3}\bigg|_{D\to C}.\label{eq:Poisson-2-in-Thetas}
\end{align}

Carrying out all the above computations we get
\begin{align}
P_{2}= & -\kappa\underset{J\leftrightarrow K}{A}\underset{I\leftrightarrow K}{S}D_{J}^{x}\left[\delta^3\left(x-y\right)R^{F}{}_{0CD}\left(\pi_o^x\right)\right]\eta_{BI}\eta_{FK}\nonumber \\
 & +\frac{\kappa}{4}\underset{J\leftrightarrow K}{A}\underset{I\leftrightarrow K}{S}D_{F}^{x}\left[\delta^3\left(x-y\right)R^{F}{}_{0CD}\left(\pi_o^x\right)\right]\eta_{BI}\eta_{JK}\nonumber \\
 & +\frac{\kappa}{4}\underset{J\leftrightarrow K}{A}\underset{I\leftrightarrow K}{S}D_{B}^{x}\left[\delta^3\left(x-y\right)R^{F}{}_{0CD}\left(\pi_o^x\right)\right]\eta_{JI}\eta_{FK}\nonumber \\
 & -\frac{\kappa}{12}\underset{C\leftrightarrow D}{A}\underset{J\leftrightarrow K}{A}\underset{I\leftrightarrow K}{S}D_{M}^{x}\left[\delta^3\left(x-y\right)\left(R^{M}{}_{JC0}\left(\pi_o^x\right)+R^{M}{}_{CJ0}\left(\pi_o^x\right)\right)\right]\eta_{BI}\eta_{DK}\nonumber \\
 & +\frac{\kappa}{12}\underset{C\leftrightarrow D}{A}\underset{J\leftrightarrow K}{A}\underset{I\leftrightarrow K}{S}D_{J}^{x}\left[\delta^3\left(x-y\right)\left(R^{M}{}_{CB0}\left(\pi_o^x\right)+R^{M}{}_{BC0}\left(\pi_o^x\right)\right)\right]\eta_{MI}\eta_{DK}\nonumber \\
 & +\frac{\kappa}{12}\underset{C\leftrightarrow D}{A}\underset{J\leftrightarrow K}{A}\underset{I\leftrightarrow K}{S}D_{B}^{x}\left[\delta^3\left(x-y\right)\left(R^{M}{}_{JC0}\left(\pi_o^x\right)+R^{M}{}_{CJ0}\left(\pi_o^x\right)\right)\right]\eta_{MI}\eta_{DK}\nonumber \\
 & +\frac{\kappa}{12}\underset{C\leftrightarrow D}{A}\underset{J\leftrightarrow K}{A}\underset{I\leftrightarrow K}{S}D_{C}^{x}\left[\delta^3\left(x-y\right)\left(R^{M}{}_{BJ0}\left(\pi_o^x\right)+R^{M}{}_{JB0}\left(\pi_o^x\right)\right)\right]\eta_{MI}\eta_{DK}\nonumber \\
 & +\kappa\underset{C\leftrightarrow D}{A}\underset{J\leftrightarrow K}{A}\underset{I\leftrightarrow K}{S}D_{J}^{x}\left[\delta^3\left(x-y\right)R^{F}{}_{CB0}\left(\pi_o^x\right)\right]\eta_{DI}\eta_{FK}\nonumber \\
 & -\frac{\kappa}{4}\underset{C\leftrightarrow D}{A}\underset{J\leftrightarrow K}{A}\underset{I\leftrightarrow K}{S}D_{F}^{x}\left[\delta^3\left(x-y\right)R^{F}{}_{CB0}\left(\pi_o^x\right)\right]\eta_{DI}\eta_{JK}\nonumber \\
 & -\frac{\kappa}{4}\underset{C\leftrightarrow D}{A}\underset{J\leftrightarrow K}{A}\underset{I\leftrightarrow K}{S}D_{D}^{x}\left[\delta^3\left(x-y\right)R^{F}{}_{CB0}\left(\pi_o^x\right)\right]\eta_{JI}\eta_{FK}\nonumber \\
 & -\frac{\kappa}{12}\underset{C\leftrightarrow D}{A}\underset{J\leftrightarrow K}{A}\underset{I\leftrightarrow K}{S}D_{J}^{x}\left[\delta^3\left(x-y\right)\left(R^{M}{}_{0BD}\left(\pi_o^x\right)+R^{M}{}_{DB0}\left(\pi_o^x\right)\right)\right]\eta_{MI}\eta_{CK}\nonumber \\
 & -\frac{\kappa}{12}\underset{C\leftrightarrow D}{A}\underset{J\leftrightarrow K}{A}\underset{I\leftrightarrow K}{S}D_{J}^{x}\left[\delta^3\left(x-y\right)\left(R^{M}{}_{0CD}\left(\pi_o^x\right)+R^{M}{}_{DC0}\left(\pi_o^x\right)\right)\right]\eta_{BI}\eta_{MK}\nonumber \\
 & +\frac{\kappa}{12}\underset{C\leftrightarrow D}{A}\underset{J\leftrightarrow K}{A}\underset{I\leftrightarrow K}{S}D_{J}^{x}\left[\delta^3\left(x-y\right)\left(R_{BDC0}\left(\pi_o^x\right)+R_{CDB0}\left(\pi_o^x\right)\right)\right]\eta_{IK}\nonumber \\
 & -\frac{\kappa}{12}\underset{C\leftrightarrow D}{A}\underset{J\leftrightarrow K}{A}\underset{I\leftrightarrow K}{S}D_{F}^{x}\left[\delta^3\left(x-y\right)\left(R_{JDB0}\left(\pi_o^x\right)+R_{BDJ0}\left(\pi_o^x\right)\right)\right]\eta^{MF}\eta_{MI}\eta_{CK}\nonumber \\
 & +\frac{\kappa}{6}\underset{C\leftrightarrow D}{A}\underset{J\leftrightarrow K}{A}\underset{I\leftrightarrow K}{S}D_{F}^{x}\left[\delta^3\left(x-y\right)\left(R^{F}{}_{DJ0}\left(\pi_o^x\right)+R^{F}{}_{0JD}\left(\pi_o^x\right)\right)\right]\eta_{BI}\eta_{CK}\nonumber \\
 & -\frac{\kappa}{12}\underset{C\leftrightarrow D}{A}\underset{J\leftrightarrow K}{A}\underset{I\leftrightarrow K}{S}D_{F}^{x}\left[\delta^3\left(x-y\right)\left(R_{JDC0}\left(\pi_o^x\right)+R_{CDJ0}\left(\pi_o^x\right)\right)\right]\eta^{MF}\eta_{BI}\eta_{MK}\nonumber \\
 & -\frac{\kappa}{12}\underset{C\leftrightarrow D}{A}\underset{J\leftrightarrow K}{A}\underset{I\leftrightarrow K}{S}D_{B}^{x}\left[\delta^3\left(x-y\right)\left(R^{M}{}_{DJ0}\left(\pi_o^x\right)+R^{M}{}_{0JD}\left(\pi_o^x\right)\right)\right]\eta_{MI}\eta_{CK}\nonumber \\
 & +\frac{\kappa}{12}\underset{C\leftrightarrow D}{A}\underset{J\leftrightarrow K}{A}\underset{I\leftrightarrow K}{S}D_{B}^{x}\left[\delta^3\left(x-y\right)\left(R_{CDJ0}\left(\pi_o^x\right)+R_{JDC0}\left(\pi_o^x\right)\right)\right]\eta_{IK}\nonumber \\
 & -\frac{\kappa}{12}\underset{C\leftrightarrow D}{A}\underset{J\leftrightarrow K}{A}\underset{I\leftrightarrow K}{S}D_{C}^{x}\left[\delta^3\left(x-y\right)\left(R^{M}{}_{DJ0}\left(\pi_o^x\right)+R^{M}{}_{0JD}\left(\pi_o^x\right)\right)\right]\eta_{BI}\eta_{MK}\nonumber \\
 & +\frac{\kappa}{12}\underset{C\leftrightarrow
   D}{A}\underset{J\leftrightarrow K}{A}\underset{I\leftrightarrow
   K}{S}D_{C}^{x}\left[\delta^3\left(x-y\right)\left(R_{JDB0}\left(\pi_o^x\right)+R_{BDJ0}\left(\pi_o^x\right)
   \right)\right]\eta_{IK}.
\end{align}
and as before there is a consistency relation with the equations of
motion,
\begin{equation}
\eta^{AB}  \left\{ R_{0ABC}\left(\pi_o^x\right),R_{0IJK}\left(\pi_o^x\circ  \pi_x^y\right)\right\}=0.
\end{equation}
And the expression can also be made more compact through the
symmetrizer, as before,
\begin{align}
P_{2}= & -\kappa\underset{J\leftrightarrow K}{A}\underset{I\leftrightarrow K}{S}D_{J}^{x}\left[\phi\left(x\right)R^{F}{}_{0CD}\left(x\right)\right]\eta_{BI}\eta_{FK}\nonumber \\
 & -\kappa\underset{C\leftrightarrow D}{A}\underset{J\leftrightarrow K}{A}\underset{I\leftrightarrow K}{S}D_{J}^{x}\left[\phi\left(x\right)R^{F}{}_{CB0}\left(x\right)\right]\eta_{DI}\eta_{FK}\nonumber \\
 & +\frac{\kappa}{4}\underset{J\leftrightarrow K}{A}\underset{I\leftrightarrow K}{S}\underset{F\leftrightarrow B}{T}D_{F}^{x}\left[\phi\left(x\right)R^{F}{}_{0CD}\left(x\right)\right]\eta_{BI}\eta_{JK}\nonumber \\
 & +\frac{\kappa}{4}\underset{C\leftrightarrow D}{A}\underset{J\leftrightarrow K}{A}\underset{I\leftrightarrow K}{S}\underset{F\leftrightarrow D}{T}D_{F}^{x}\left[\phi\left(x\right)R^{F}{}_{CB0}\left(x\right)\right]\eta_{IJ}\eta_{DK}\nonumber \\
 & -\frac{\kappa}{12}\underset{C\leftrightarrow D}{A}\underset{J\leftrightarrow K}{A}\underset{I\leftrightarrow K}{S}\underset{J\leftrightarrow C}{T}\underset{F\leftrightarrow B}{A}D_{F}^{x}\left[\phi\left(x\right)R^{F}{}_{JC0}\left(x\right)\right]\eta_{BI}\eta_{DK}\nonumber \\
 & +\frac{\kappa}{12}\underset{C\leftrightarrow D}{A}\underset{J\leftrightarrow K}{A}\underset{I\leftrightarrow K}{S}\underset{B\leftrightarrow C}{T}\underset{J\leftrightarrow C}{T}D_{J}^{x}\left[\phi\left(x\right)R^{F}{}_{BC0}\left(x\right)\right]\eta_{FI}\eta_{DK}\nonumber \\
 & -\frac{\kappa}{12}\underset{C\leftrightarrow D}{A}\underset{J\leftrightarrow K}{A}\underset{I\leftrightarrow K}{S}\underset{I\leftrightarrow C}{A}\underset{I\leftrightarrow B}{T}\underset{J\leftrightarrow B}{T}D_{I}^{x}\left[\phi\left(x\right)R_{0BDJ}\left(x\right)\right]\eta_{CK}\nonumber \\
 & -\frac{\kappa}{12}\underset{C\leftrightarrow D}{A}\underset{J\leftrightarrow K}{A}\underset{I\leftrightarrow K}{S}\underset{I\leftrightarrow C}{A}\underset{I\leftrightarrow B}{T}D_{J}^{x}\left[\phi\left(x\right)R_{0IDB}\left(x\right)\right]\eta_{CK}\nonumber \\
 & -\frac{\kappa}{12}\underset{C\leftrightarrow D}{A}\underset{J\leftrightarrow K}{A}\underset{I\leftrightarrow K}{S}\underset{K\leftrightarrow J}{T}\underset{C\leftrightarrow J}{T}D_{K}^{x}\left[\phi\left(x\right)R_{0CDJ}\left(x\right)\right]\eta_{IB}\nonumber \\
 & -\frac{\kappa}{12}\underset{C\leftrightarrow D}{A}\underset{J\leftrightarrow K}{A}\underset{I\leftrightarrow K}{S}\underset{K\leftrightarrow J}{T}D_{C}^{x}\left[\phi\left(x\right)R_{0KDJ}\left(x\right)\right]\eta_{BI}\nonumber \\
 & +\frac{\kappa}{6}\underset{C\leftrightarrow D}{A}\underset{J\leftrightarrow K}{A}\underset{I\leftrightarrow K}{S}\underset{0\leftrightarrow D}{T}D_{F}^{x}\left[\phi\left(x\right)R^{F}{}_{0JD}\left(x\right)\right]\eta_{BI}\eta_{CK}\nonumber \\
 & -\frac{\kappa}{12}\underset{C\leftrightarrow D}{A}\underset{J\leftrightarrow K}{A}\underset{I\leftrightarrow K}{S}\underset{J\leftrightarrow C}{T}D_{J}^{x}\left[\phi\left(x\right)R_{C0}\left(x\right)\right]\eta_{BI}\eta_{DK}\nonumber \\
 & +\frac{\kappa}{6}\underset{C\leftrightarrow D}{A}\underset{J\leftrightarrow K}{A}\underset{I\leftrightarrow K}{S}D_{J}^{x}\left[\phi\left(x\right)R_{D0}\left(x\right)\right]\eta_{BI}\eta_{CK}.\label{eq:full-final-PB-2}
\end{align}

The remaining Poisson brackets are pretty straightforward to compute.
Some of them vanish,
\begin{eqnarray}
  \label{eq:5}
  \left\{
  R_{0A0C}\left(\pi_o^x\right),R_{0I0K}\left(\pi_o^x\circ\pi_x^y\right)\right\}&=&0,\\
  \left\{  R_{0A0C}\left(\pi_o^x\right),R_{IJKL}\left(\pi_o^x\circ\pi_x^y\right)\right\}&=&0,
\end{eqnarray}
and the others can be written in terms of $P_{1,2}$ using the
equations of motion and are non-vanishing,
\begin{equation}
  \label{eq:6}
  \left\{
    R_{0A0C}\left(\pi_o^x\right),R_{0IJK}\left(\pi_o^x\circ\pi_x^y\right)\right\}=  \left\{
    R_{BABC}\left(\pi_o^x\right),R_{0IJK}\left(\pi_o^x\circ\pi_x^y\right)\right\}.
\end{equation}

It is possible to extend the algebra to arbitrary paths using the
deformation techniques here developed. The analysis must be extended
to paths in any spatial or time-like direction,
\begin{equation}
  \left\{
    R_{abcd}\left(\pi_o^x\right),R_{ijkl}\left(\eta_o^y\right)\right\}=  
\Delta_{abcd,ijkl}\left(\pi_o^x,\eta_o^y,[R]\right),
\end{equation}
Where $\Delta$ is a path-dependent distribution that takes non vanishing values when $\pi$ and $\eta$
end on the same point for the geometry given by the intrinsic Riemann tensor $R$. 
The problem of dynamics, that up to now was unsolvable in loop quantum gravity 
arises in this context as the computation of the algebra for arbitrary paths and its 
operatorial implementation.

\section{Conclusions}
As Donnelly and Giddings \cite{GiDo} put it, and results from (5.14),
``The physical observables in a gravitational theory therefore do not organize themselves 
into local commuting sub-algebras'' [as occurs in usual field theories]: ``the principle of locality
must apparently be reformulated or abandoned, and in fact we lack a clear definition 
of the coarser  and more basic notion of a quantum subsystem of the Universe.'' 

Locality expresses the idea that quantum processes can be
localized in space and time [and, at the level of observable quantities, that causally
separated processes are exempt from any uncertainty relations restricting their
co-measurability.

The quantum implementation of the Poisson algebra of intrinsic Riemann
tensors presented in this paper could provide an approximate notion of
quantum subsystem and allow to determine uncertainty relations
restricting the co-measurability of physical observables.

The quantization of the general framework laid out in this paper is
clearly a tall order. However, it can be the starting point for the
analysis of simplified situations, like minisuperspaces. We plan on
pursuing this in the future in order to identify the fundamental
elements of a quantum version of this framework.

\section{Acknowledgments}
This work was supported in part by Grant
No. NSF-PHY-1603630, NSF-PHY-1903799, funds of the Hearne Institute for Theoretical
Physics, CCT-LSU, and Pedeciba and Fondo Clemente Estable
FCE\_1\_2019\_1\_155865.

\end{document}